\documentclass[10pt,a4paper,notitlepage]{article}
\usepackage{geometry}             
\usepackage{extsizes}
\usepackage{amssymb}
\usepackage{amsmath}
\usepackage{amsthm}
\usepackage{url}
\usepackage{placeins}
\usepackage{subfigure}
\usepackage{multirow}
\usepackage{epsfig}
\newtheorem{theorem}{Theorem}

\newtheorem{definition}{Definition}
\newcommand{\Pbb}{\ensuremath{\mathbb{P}}}

\newcommand{\Qbb}{\ensuremath{\mathbb{Q}}}

\begin{document}
\author{Chris Kenyon and Andrew Green\footnote{Contact: chris.kenyon@lloydsbanking.com}}
\title{Regulatory-Compliant Derivatives Pricing\\ is Not Risk-Neutral\footnote{\bf The views expressed are those of the authors only, no other representation should be attributed.}}
\date{\today, Version 2.20}

\maketitle

\begin{abstract}
Regulations impose idiosyncratic capital and funding costs for holding derivatives.    Capital requirements are costly because derivatives desks are risky businesses; funding is costly in part because regulations increase the minimum funding tenor.   Idiosyncratic costs mean no single measure makes derivatives martingales for all market participants.  Hence Regulatory-compliant pricing is not risk-neutral. This has implications for exit prices and mark-to-market.
\end{abstract}
 
\section{Introduction}

Increased regulations, and market changes, since 2007 have altered the perceived costs of many financial products.  Here we prove that these changes are not just perception but that they have had a fundamental effect on pricing theory.  That is, we prove that a market-wide risk-neutral measure that is common for all market participants does not exist.  This proof is based on our Theorem \ref{th:one} which states that if different market participants have different dividends for holding the same stock then there is no market-wide risk-neutral measure that is common for all market participants.  We then demonstrate that due to regulations, and unhedgable risks, different trading businesses do have different holding costs for the same positions.  This means that today all valuations are private in the sense that they can be derived from idiosyncratic risk-neutral measures (i.e. local to the individual pricing institution).  Executable screen prices are components of value, not valuations by themselves because of these idiosyncratic and asymmetric costs.

Legally binding regulations require institutions to hold capital, e.g. CRD IV in Europe \cite{CRD-IV-Regulation} and Dodd-Frank in the USA \cite{Dodd2010a}.  Regulations also contain liquidity requirements such as the Liquidity Coverage Ratio \cite{BCBS-189}.  Bilateral initial margins are proposed in \cite{BCBS-261} as well as being required by central counterparties whose use is mandated for certain product classes \cite{Dodd2010a}.  Thus, in general, there are unavoidable capital and funding requirements even for flow desks with fully collateralized back-to-back trades. 

If capital is provided at the riskless overnight rate, and liquidity used in liquidity buffers is also provided at the riskless overnight rate\footnote{We assume that a riskless rate is available to investors.  This is not strictly required here but we adopt it for convenience.}, then these regulations have no effect on pricing.  If capital and funding are not provided at the riskless overnight rate, and  different amounts are  required by different institutions for the same position, then we demonstrate that there is no market-wide risk-neutral measure that is the same for all participants.

In theory, should investors provide capital, and funding, at the riskless overnight rate to a trading business?  If the (trading) business has risk then no, and we will demonstrate in detail that trading businesses have unhedgable risks.  Furthermore, since the Basel III Liquidity Coverage Ratio \cite{BCBS-189}  has a minimum term of one month, overnight funding is not usable for funding buffers.    Thus there are at least two independent sources of additional costs above riskless to a trading business.  This appears to lead to the conclusions that trading business must continually lose money.  However, trading businesses rely on competitive advantages to make profits --- just like any other business. 

Technically this paper adds to the literature on derivatives with holding costs \cite{Tuckman1992a}, rather than transaction costs \cite{Kabanov2010a}.  It is also related to the incomplete-markets literature \cite{Cerny2009a,Kaido2009a}.

The main contribution of this paper is to show that regulations fundamentally change pricing in that there is no common, market-wide, risk-neutral measure.   A second contribution is to show how the concept of an exit prices can be understood in the absence of a market-wide risk-neutral measure.  We also address marketing-to-market and the FVA debate.
\section{No Market-Wide Risk-Neutral Measure}

We use two steps: firstly we identify a technical condition under which no common market-wide risk neutral measure exists; secondly we demonstrate this condition holds in practice.

\subsection{Condition for the Non-Existence of the Risk-Neutral Measure}

\begin{definition} [\cite{Shreve2004a}] Let \Pbb\ be the physical measure, then a probability measure \Qbb\ is said to be risk-neutral if:

(i) \Qbb\ and \Pbb\ are equivalent;

(ii) under \Qbb\ discounted stock prices are martingales.

\end{definition}

\noindent
\cite{Shreve2004a} shows that if the market price of risk equations cannot be solved then there is arbitrage assuming that the cost of 1 unit of stock is exactly the negative of -1 units of stock.  In our case this is not true because holding costs are present whether the position is short or long.

\begin{theorem}\label{th:one}
If there are market participant with different idiosyncratic continuous dividends when holding the same stock then there is no common risk-neutral measure for all participants. \label{th:no}
\end{theorem}
\begin{proof}
Obvious. Let the stock price, from the point of view of market participant $i$, be:
\[
dS_i(t) = (\mu_i + a_i)S_i(t) dt + \sigma S_i(t) dW^{\Pbb_i}(t)
\]
where $a_i$ is the objective dividend received by market participant $i$, and $\mu_i$ is the \Pbb\ drift believed by market participant $i$.  This implies that in the idiosyncratic risk-neutral measure of $i$, the evolution of the stock price is:
\[
dS_i(t) = (r + a_i) S_i(t) dt + \sigma S_i(t) dW^{\Qbb_i}(t)
\]
where $r$ is the riskless rate.  The \Pbb\  drifts of the market participants have been replaced by the riskless rate, but dividends are unchanged because they are objective although idiosyncratic.  Hence there is no common risk neutral measure for all participants because the rates of return are different participants with different dividends (under each participants' risk neutral measures).  
\end{proof}
If dividends were not idiosyncratic then all participants would see the same risk-neutral measure.  Usually the Girsanov transformations are idiosyncratic but the final measure is common.

The condition for the non-existence of the risk-neutral measure, is that there are participants with different holding costs (negative dividend) for holding the same stock.  It is clear that the theorem applies to any self-financing portfolio, i.e. any derivative.  The condition is not long and technical, rather it is simple and direct.

\subsection{Existence of the Condition}\label{s:exist}
Here we show that different market participants have different, i.e. idiosyncratic, holding costs for the same derivative.  We do this in two steps: firstly we show that different participants have different capital and funding requirements for the same derivative. Secondly we show that capital and funding have non-zero costs for market participants.  

If capital and funding have different units costs for different market participants then the conditions of the theorem are also fulfilled\footnote{Unless these cancel out precisely, which is vanishingly unlikely over the whole market at all times.}.  However different unit costs are not required because we show that different quantities of funding and capital can be required for the same positions.

\subsubsection{Different capital and funding requirements for the same derivative}

We can pick any derivative for this section\footnote{We can pick any derivative because we are constructing a counter-example to the existence of a market-wide risk-neutral measure that is common for all participants.}, and we choose interest rate swaps as one of the most liquid derivatives.  For concreteness we take the swaps as being EUR currency and that the clients are in the Eurozone.  Consider the following setups. 

\begin{figure}[htbp]
	\centering
		\includegraphics[width=1.1\textwidth,clip,trim=60 50 110 0]{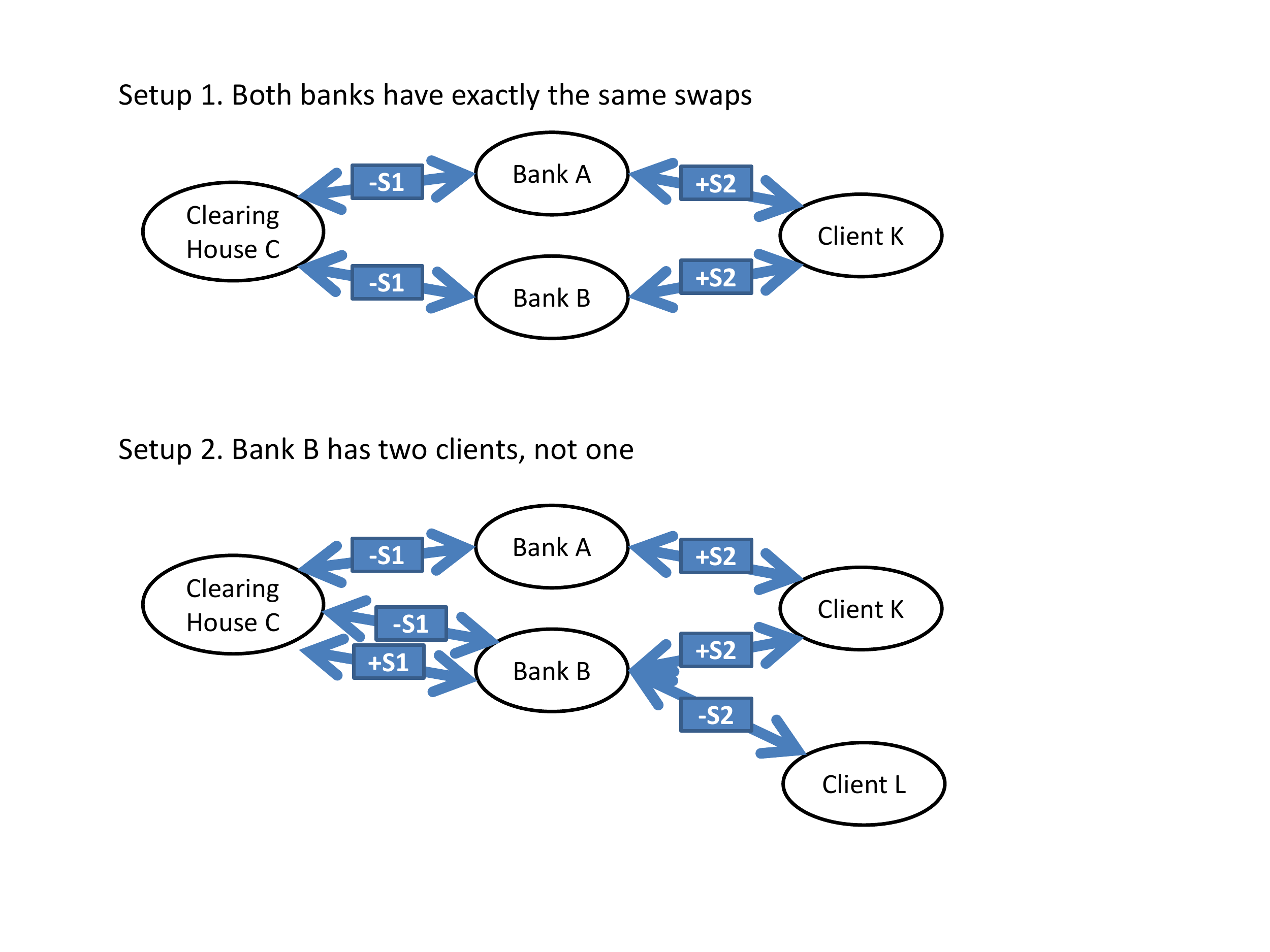}
	\caption{The two setups described in Section \ref{s:exist}.  $\pm S1$ and $\pm S2$ stand for the swap positions held by the different counterparties.  The sign gives the direction.  The details of the swaps $S1$ and $S2$ (ignoring the sign) are identical apart from the swap rate.  The client swaps ($S2$) are uncollateralized so the swap rate takes into account higher counterparty risk than with the collateralized swaps with the clearing house ($S1$).}
	\label{f:counterparties}
\end{figure}

\paragraph{Setup 1. Both banks have exactly the same trades}  This is shown in the top half of Figure \ref{f:counterparties}.
\begin{itemize}
	\item Bank A has two back-to-back interest rate swaps, one with clearing house C, and one with a non-financial client K.  The swap with the clearing house is collateralized, the swap with the client is not.  
	\item Bank B has exactly the same setup as Bank A (i.e. trades with the same client and clearing house).
\end{itemize}
The two banks will have different spot, and lifetime, regulatory capital requirements in at least the following cases.  This list is not exhaustive.
\begin{enumerate}
	\item Bank A has Regulatory permission to use Internal Models for VaR and bank B does not.
	\item Bank A is a Systematically Important Bank and bank B is not (because systematically important banks have additional capital requirements).
	\item Banks A and B are both Systematically Important, but with different capital add-ons.
	\item Bank A is European and the non-financial client is small enough that the CRD IV exemption for CVA VaR for non-financial entities is applicable\footnote{See Article 381, point 4, of Regulation (EU) No 575/2013 amending Regulation (EU) 648/2012.}, and bank B is non-European where the CVA VaR charge has no non-financial client exemptions.
\end{enumerate}
Capital requirements include: Counterparty Credit Risk capital; CVA VaR capital; Market Risk capital; Default Fund capital.  There will be Market Risk on all the swaps because the swap rate on the client side will not be substantially similar (i.e. within 15bps) to that on the clearing house side.  The swap rates will be different because the swap rate on client side will price in the possibility of counterparty default (probably much greater than 15bps).  The default fund capital charge is not yet a regulatory requirement, but it is sufficiently close that both banks will price it into their lifetime capital costs.
Both banks will have funding requirements, but these will be identical.

\paragraph{Setup 2. Bank B has two clients, not one} This is shown in the bottom half of Figure \ref{f:counterparties}.
\begin{itemize}
	\item Bank B has identical back-to-back interest rate swaps, one with clearing house C, and one with another non-financial client K.  In addition, bank B has two more back-to-back interest rate swaps, another with the clearing house, and the final one with a second non-financial client, L which has the same credit characteristics as K.  These further swaps are in opposite directions to the first pair, but equal notionals.  Hence bank B has offsetting swaps with clearing house C.  All trades with the clearing house are collateralized. 
\end{itemize}
Considering the back-to-back swaps involving client K, capital and funding requirements diverge further between the two banks than in the previous setup.

We assume that bank B attributes funding and capital costs to each pair of back-to-back swaps using a marginal-effect procedure.  A marginal-effect procedure is where each item is removed to see its effect, and then the totals are normalized.  In this case an item would be a pair of back-to-back swaps.
\begin{itemize}
	\item Funding requirements will be different for the back-to-back swaps with client K between bank A and bank B, because (this list is not exhaustive):
	\begin{itemize}
		\item Initial Margin at the clearing house will be significantly lower for bank B than for bank A.
		\item Variation Margin at the clearing house will be zero for bank B and non-zero for bank A.
		\item Default Fund contributions will be relatively different for the two banks, because of the netting of the two swaps with C for Bank B.  We assume the clearing house requires default fund contributions from the banks according to their Initial Margins normalized with respect to all clearing members.  There are more complex schemes but we make this assumption for simplicity.
	\end{itemize}
	\item Capital requirements will be different for the back-to-back swaps with client K between bank A and bank B (in addition to the reasons in the first setup) because:
	\begin{itemize}
		\item Bank B has no market risk for the swaps at the clearing house because they net exactly.
		\item The two banks have different Default Fund capital requirements from their different default fund contributions.
		
		This list is not exhaustive.
	\end{itemize}
\end{itemize}
The presence of the addition pair of back-to-back swaps significantly alters funding and capital requirements for bank B.  This is not just a change in the absolute levels but also a change in the {\it attributable} levels for each pair of back-to-back swaps.

\paragraph{Setup 3. Realistic portfolios}

The divergence of capital and funding requirements, for the same derivatives, will hold for real portfolios.  Banks have different regulatory status, and varying portfolios with different netting benefits.  Banks may also have different stress periods for their Stressed VaR in Market Risk.

Even assuming banks with indistinguishable risk levels, their portfolios and regulatory status will produce different capital and funding requirements.  However, we deal with risk levels next.

\FloatBarrier
\subsubsection{Capital and Funding are not Priced as Riskless to Trading Businesses in Theory}
This is the second theoretical step required to prove the non-existence of a common market-wide risk-neutral measure.  Note that this is a theoretical claim, we do not depend on whatever Treasury does --- or does not --- charge to trading businesses in practice.

This step is required.  It is not sufficient to show that different trading businesses require different quantities of capital and funding for the same derivative.  We must also show that capital and funding have non-zero costs.  Otherwise these requirements do not matter.  

\paragraph{To prove that capital and funding are priced above riskless, we must show that trading businesses are risky.}
We can show that trading businesses are risky by showing that they have unavoidable PnL\footnote{Profit-and-Loss.} leaks.  There are several basic PnL leaks.
\begin{enumerate}
	\item All desks leak their institutional costs (IC), i.e. salaries and facilities.  Consider a flow swaps desk with only collateralized trades.  If all its trades are at-the-money (ATM) the desk makes losses the size of its IC.  To pay its IC it must be largely on the good side of the bid-ask spread.  How can a desk be consistently on the good side of bid-ask spreads?  This is a consequence of the business model of the desk.  Now all business models have some risk, hence the desk is risky.  
	\item When there is a funding requirement (e.g. Initial Margins) Regulatory funding buffers are required and these create PnL leaks.  The buffers increase funding costs, per unit of funding, by changing the minimum funding maturity.  Funding can be optimized, \cite{Kenyon2013c2}, but the cost remains.  
\end{enumerate}
The flow swaps desk in the example above may be one of the least risky business models, but there is still some risk.  Hence its funding (Initial Margins at least) and capital (probably some Market Risk, Default Fund capital, and credit risk on margin period of risk) will have costs above riskless from investors.  However, the capital and funding quantities can only generate the riskless rate of return.  Thus the funding and capital requirements multiply the risk of the desk.   

It can be argued that the capital and funding requirements on the desk make the desk safer.  However, capital and funding make the desk safer from market, credit, and funding risk --- but they do not patch PnL leaks.  The above-riskless cost of funding and capital multiplies the PnL leaks and puts increased stress on the business model of the desk, thus making it riskier.

We do not expect theoretical investors to perceive any derivatives business as riskless.  The institutional costs PnL leak of the swaps desk was simply a basic example.  Away from flow desks, business risks can increase, for example:
\begin{itemize}
	\item Any dynamic hedging strategy.  The risks here come from the finite hedging interval and the requirement that the hedging instruments remain liquid.  Over the crisis the liquidity in many markets changed and significant gap events were observed.  
	\item Any insurance-type contract, e.g. a CDS (credit default swap).  Consider the case where the reference entity defaults one day after protection was bought.  The protection-buyer makes a profit.   Hence credit hedges do not hedge in every state of the world.  This is a significant difference relative to, say, interest rate hedging.
	\item Any incomplete market.  Depending on how the market is incomplete and the product there may be unbounded risks and requirements for dynamic hedging.
	\item Costly changes in regulations.  Regulations are generally well-announced, but their exact timing and contents are often uncertain.
\end{itemize}
These unhedgable risks, together with institutional costs, mean that trading desks are risky businesses even in theory.  Thus they will be charged above riskless for their funding and capital by investors.

We have now demonstrated that the condition required in Theorem 1 holds and there is no common market-wide risk-neutral measure for all participants.

\section{Implications}

We consider exit prices, marking to market, and the FVA debate \cite{Hull2013a,Carver2012b}.

\subsection{Exit Prices}

Exit prices are required for many purposes and often defined as fair value, or in accordance with an ISDA protocol \cite{ISDA-2009-Clo}.  The non-existence of a common risk-neutral measure for all participants is not an issue because the exit price is of most interest to the two institutions involved with the position.  The price that preserves competitive advantage is an obvious definition.  We call this flat-idiosyncratic-exit.  That is, the institution with the competitive advantage is flat (no profit and no loss considering all costs).  This is an idiosyncratic valuation, but we have proved that all valuations must be idiosyncratic valuations.
There are two cases of interest: defaulted counterparty and existing counterparty.  Clearly the competitive advantage lies with the survivor in the case where one party has defaulted.  For existing counterparties the competitive advantage generally lies with the party that does not initiate the trade exit.

In Accounting, Fair Value includes the idea of willing participants (i.e. specifically not a fire-sale).  Flat-idiosyncratic-exit ensures that the non-initiating party is willing (no PnL effect) and that the initiator is also willing because they were aware of the consequential loss due to their initiation of the trade exit process.

\subsection{Marking to Market}

If all valuations are idiosyncratic, what does marketing to market mean?  This is a loaded question because, for example, an uncollateralized trade builds in the risk of both counterparties and thus is not fungible.  ``uncollateralized market price'' is an oxymoron.  Any market involving such trades can only ever involve the original counterparties.    Thus there are no market prices for uncollateralized trades because there is no market.

Marking to market collateralized and cleared trades is also problematic.  Few cleared trades are standalone --- what would be the business justification for such an open position?  Thus whilst the market level of swaps on a clearing house (e.g. LCH) is observable, the trades only exist because they are part of some package or hedging strategy.  The clearing house side of such strategies does have visible prices but these are only one component of a portfolio or netting set value.  Since clearing houses require posting of initial margin this cost, and capital costs, will be included in any trading strategy involving cleared trades.   Once bilateral initial margin comes in \cite{BCBS-261} these considerations will be universal.  Thus executable screen prices are part of valuation but only part.
  
Marking to market is an idiosyncratic exercise.  Each market participant has competitive advantages that dictate its business strategy with respect to the market and hence its realizable valuations.  That is, marking to market must take into account the relative positions of the market players, there is no homogeneity.  Thus idiosyncratic prices are not inconsistent with the market, instead they are exactly the expression of its heterogeneity.

Markets have observable prices, e.g. executable screen prices.  However, the properties of markets are not of central importance for market participants given the non-existence of a common risk-neutral measure.  What matters to each participant are the replication prices that the participant can construct using market prices as one input.  Other important inputs are the participant's existing portfolio together with funding and capital requirements from Regulations.  Each market participant will derive different replication prices from the same observed market.

\subsection{The FVA Debate}

\begin{figure}[htbp]
	\centering
		\includegraphics[width=0.80\textwidth,clip,trim=0 50 0 100]{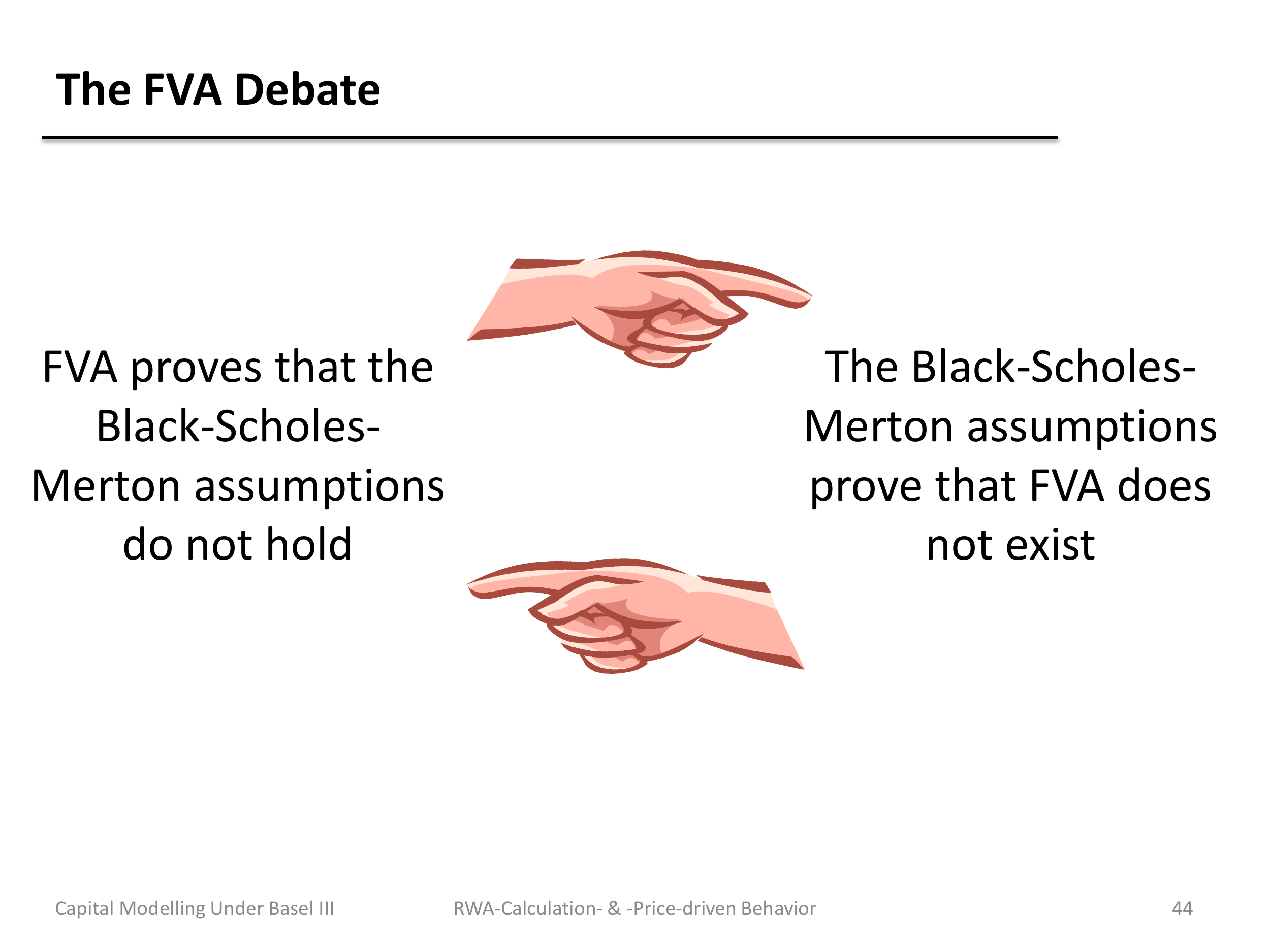}
	\caption{A view of the FVA debate.  Funding costs are part of the second step in our development showing that there is no market-wide risk-neutral measure shared by all participants.}
	\label{f:fvaDebate}
\end{figure}

Our view of this is summarized in Figure \ref{f:fvaDebate}.  Funding costs are part of the second step in our development showing that there is no market-wide risk-neutral measure.  This debate was started by \cite{Hull2012a} and continued in \cite{Hull2013a}.  Essentially the authors argue that ``the evaluation of an investment should depend on the risk of an investment ...''.  Most authors agree with this statement, the disagreements occur when applying the statement in practice.  The original authors add a further phrase ``... , not how it is financed'', and it is at this point that the debate starts.  We note that a Nobel prize was awarded for the Modilgiani-Miller Propositions \cite{Miller1988a}, which is an early replication, and makes the same statement.

Our Theorem \ref{th:one} describes a condition under which replication by individual market participants will not create unique derivatives values for market participants.  We have also demonstrated that this condition holds in practice.  Ironically, regulations that safeguard businesses from Market, Credit, and Liquidity risks, do not block PnL leaks from the costs of the safeguards. 
 
Regulations do not provide guaranteed profits.  Profits are the domain of the business model, and competitive advantage is required.  Clearly, a competitive advantage is a risk because there is no way to guarantee its future.  Competitive advantages have a lifecycle.

Modilgiani-Miller state that debt and equity investors divide the profits (or losses) of a risky project between them.  Since Modilgiani-Miller assume that the investors do not change the project's cashflows, or its risk, then the value of the project cannot be affected by the investor mix (assuming an efficient market).  We have demonstrated, as part of our development, that the firm holding the project  does change the cashflows of the project, e.g. by the Regulatory status of the firm.  It is not only the Regulatory status of the firm that will change the cashflows of the project. The firm will also change the cashflows of its projects when it adds new projects, for example due to netting with clearing houses (as in our Setup 2.).  Note that the changes in projects cashflows involve attribution of costs, for example because of netting or capital.  The changes in cashflows, of course, also change funding requirements, which changes risk (at least in as much as the funding requirements are volatile, and are not term funded).  Hence \cite{Hull2012a,Hull2013a} are incorrect because the assumptions of Modilgiani-Miller do not hold. 

It may appear that our Theorem \ref{th:one} implies that there are widespread arbitrage opportunities.  In fact, just the existence of different Regulatory regimes implies differences in costs for the same positions.  This is not theory but Law.  In theory too, holding costs can create arbitrage opportunities \cite{Tuckman1992a}.  The fact that arbitrage potentially exists in theory does not mean that it is exploitable.  We leave it for empirical studies to quantify how much arbitrage this leads to in practice.

\FloatBarrier
\section{Discussion and Conclusions}

Regulators are clearly not risk neutral, i.e. having linear and identical preferences either side of break-even for derivatives.  In practice the regulations on capital and funding in Basel III, together with the inherent riskiness of derivatives businesses, have imposed idiosyncratic holding costs for derivatives on market participants.  These idiosyncratic holding costs mean that there is no market-wide risk-neutral measure that is common for all participants.  Instead each participant has their own idiosyncratic risk-neutral measure (or measures plural in as much as a market is incomplete).  Thus regulation has fundamentally changed pricing theory.  Exit prices can be approached with this new understanding, i.e. as idiosyncratic-flat prices.  Marking to market uncollateralized trades is problematic precisely because of the non-existence of a common market-wide risk-neutral measure.

\subsection*{Acknowledgements}
We acknowledge many illuminating discussions with conference participants and colleagues, especially Julian Keenan, John Slee, Chris Dennis, Matteo Rolle, and Hringur Gretarsson.  We thank two anonymous reviewers for material assistance in improving the clarity of the exposition.

\bibliographystyle{alpha}
\bibliography{kenyon_general}
\end{document}